\documentclass[aps,twocolumn,showpacs]{revtex4}
\usepackage{graphicx}
\def\pt{$p_T$}
\def\dis{distribution}
\def\cor{correlation}
\def\de{\Delta\eta}
\def\bq{\begin{eqnarray}}
\def\eq{\end{eqnarray}}

\begin{document} 
\title {Ridge Formation Induced by Jets in $pp$ Collisions at 7 TeV}
\author
 {Rudolph C. Hwa$^1$ and C.\ B.\ Yang$^{1,2}$}
\affiliation
{$^1$Institute of Theoretical Science and Department of
Physics\\ University of Oregon, Eugene, OR 97403-5203, USA\\
\bigskip
$^2$Institute of Particle Physics, Hua-Zhong Normal
University, Wuhan 430079, P.\ R.\ China}
\date{\today}
\begin{abstract} 
An interpretation of the ridge phenomenon found in $pp$ collisions at 7 TeV  is given in terms of enhancement of soft partons due to energy loss of semihard jets. A description of ridge formation in nuclear collisions can directly be extended to $pp$ collisions, since hydrodynamics is not used, and azimuthal anisotropy is generated by semihard scattering. The observed ridge structure is then understood as a manifestation of soft-soft transverse correlation induced by semihard partons without long-range longitudinal correlation. Both the \pt\ and multiplicity dependencies are well reproduced. Some predictions are made about other observables.

\end{abstract}
\maketitle

\section{Introduction}

The observation of ridge structure in two-particle correlation in $pp$ collisions at 7 TeV by the CMS Collaboration at Large Hadron Collider (LHC) \cite{cms} has opened up the question of whether it has a similar origin as that already found at Relativistic Heavy-Ion Collider (RHIC) in Au-Au collisions at 0.2 TeV \cite{ja,bia,ba,bia1,md}.  A great deal is known about the ridge in heavy-ion collisions, since various experiments have studied two-particle (with or without trigger) and three-particle correlations.  The dominant theme is that the ridge exhibits the effect of high or intermediate-$p_T$ jets on a dense medium.  If the phenomenon seen at LHC reveals similar features upon further investigation, it would imply that soft partons of high density can be created in $pp$ collisions and can affect the passage of hard partons through them.  If not, a new mechanism needs to be found.  Various theoretical speculations have been advanced with varying degrees of attention to the specifics of the CMS data \cite{es,ad,pb}.  In this article we propose a model that is an extension of  our past interpretation of the ridge phenomena in the RHIC data, but is particularly suitable for  $pp$ collisions at LHC, since the dynamical origin is jet production rather than hydrodynamics.  We have a simple formula that can reproduce the CMS data quantitatively with the use of two parameters that can clearly describe the physics involved.

The most direct approach to the study of ridges is to consider only events selected by triggers with $p_T^{\rm trig}$ in an intermediate $p_T$ range, as first reported by Putschke \cite{jp,bia}.  The dependence of the ridge yield on centrality in nuclear collisions  indicates that the ridge is formed when there is a jet in a dense medium.   Having an exponential behavior in $p_T^{\rm assoc}$ at values less than $p_T^{\rm trig}$ suggests that the ridge particles are related to the soft partons, but they have an inverse slope larger than that of the inclusive distribution, implying an enhancement effect of the jet \cite{bia,rh}.  If triggers are not used as in the study of autocorrelation,  ridges are also observed at $|\Delta\eta|>1$ in central collisions \cite{ja,md}.  For $pp$ collisions at LHC we cannot presume the existence of a dense medium of partons, which is a possibility we leave open.  However, we can and shall assume that ridge formation is due to high- or intermediate-$p_T$ jets, whether or not the jets are detected by triggers.      Our goal is to study the properties of  correlation generated by semihard jets.  It should be noted that there are models in which the ridge phenomenon can occur without jets, such as in Refs. \cite{pb,gmm,ad1,kd,ar}.

In the hadronization model based on Refs.\ \cite{hy,rh}  the ridge component (due to the recombination of thermal partons) manifests the effect of the semihard parton on the  medium.  The soft partons  have  exponential dependence on the transverse momentum $k_T$, whose  inverse slope is  $T$ in the absence of semihard partons. For the ridge component the inverse slope is increased to $T'>T$ due to the enhancement of the thermal motion of the soft partons caused by the energy loss of the semihard parton that passes through the medium in the vicinity \cite{chy}. That is  soft-semihard correlation, which we shall apply to even $pp$ collisions where the notion of thermal partons may be questionable.  It is known empirically that there exists an exponential peak at small $p_T$ at LHC \cite{ga,vk,ka}; that is sufficient for us to refer to the underlying partons as soft, the recombination of which gives the low-$p_T$ hadrons.

	 In Sec.\ II  we give a short summary of our past work on ridges with emphasis on the distinction between transverse and longitudinal correlations. It is significant to note that the data on ridge reported by PHOBOS \cite{ba} do not imply the existence of long-range longitudinal \cor\ upon closer examination. In Sec.\ III the transverse \cor\ is extended to $|\de|>1$ appropriate for CMS measurement. Quantitative analysis of the ridge yield in $pp$ collisions is then carried out in Sec.\ IV. In the last section we give the conclusion along with some predictions.

\section{Transverse and Longitudinal  Correlations}

Longitudinal correlation has been the primary concern of most theoretical studies on ridges \cite{gmm,ad1,kd,cyw,mg,kw}. The observation by PHOBOS \cite{ba} that $|\Delta\eta|$ can be as large as 4 has led to the conclusion that there is empirical evidence for long-range correlation, which is an inherent property of flux-tube models. 
There are, however, two other aspects about the ridge structure that one should also consider in addition to the large-$\de$ aspect of the PHOBOS data. One is {\it A}: the property of ridge in the small $\de$ limit, and the other is {\it B}: the question of how large should $\de$ be in order for the \cor\ to be regarded as long-range. We comment on them in the context of what have been observed at RHIC as a prelude to our discussion about the ridge found at LHC.

\noindent {\it A. Transverse Correlation.} At midrapidity dihadron correlations in the azimuthal angles  have been studied in detail at RHIC; in particular, the dependence of the azimuthal \cor\ on the trigger angle $\phi_s$ relative to the reaction plane reveals features that are important about ridge formation \cite{af,dn,ma,ha}.  Any model on the origin of ridges at $|\Delta\eta|>1$ should contain properties that are consistent with the azimuthal behavior at $|\Delta\eta|<1$, since all observed ridge structure have common behavior in $\Delta\phi$ throughout the $\de$ range. The ridge yield as a function of $\phi_s$ has been studied in a model where the angular \cor\ between the trigger and local flow direction is limited \cite{ch}. It is found that a Gaussian width of $\sigma \sim 0.34$ can reproduce the data \cite{af,ma,ha}. 
The model suggests that thermal activities of the soft partons in the vicinity of the trajectory of the semihard parton (i.e., within a cone of angular range of $\sigma$) are enhanced by the energy loss of the latter to the medium. Those enhanced thermal partons hadronize into the ridge particles that rise above the background. That is transverse \cor\ between the soft and semihard partons, the only type that can be studied when $|\Delta\eta|$ is restricted to $<1$. After finding satisfactory explanation of the azimuthal \cor\ in the data this way for triggered events, the natural question to follow is how such \cor\ influences the single-particle \dis\ when triggers are not used. Semihard partons can be pervasive if their $k_T$ is around 3 GeV/c or lower. It is found that the semihard-soft transverse \cor\ can give rise to a significant azimuthal anisotropy \cite{chy,hz}, and that $v_2(p_T,N_{\rm part})$ can be quantitatively reproduced as a consequence of the ridge effect in inclusive \dis\ \cite{hz1}. This will become a key input in our discussion below where the nature of the transverse \cor\ will be made explicit.
 
\noindent {\it B. Longitudinal Correlation.}  At first sight of the PHOBOS data on the $\Delta\eta$ range of the ridge \dis\ \cite{ba}, anyone having some familiarity with multiparticle production is likely to regard $|\eta_2-\eta_1|\sim 4$ as indicative of long-range \cor\ between the trigger at $\eta_1$ and ridge particle at $\eta_2$. However, to quantify the notion of correlation range it is important to compare it to the $\eta$-range of the single-particle \dis.
 A recent study shows that the ridge \dis\ in $\de$, denoted by $dN_R^{ch}/d\Delta\eta$, can be related empirically to the single-particle \dis, $dN^{ch}/d\eta$, by using the two relevant sets of PHOBOS data only \cite{ba, bbb} without any theoretical input \cite{ch1}. That phenomenological relationship
 \bq
 {dN_R^{ch}\over d\de} \propto \int_0^{1.5} d\eta_1\left. {dN^{ch}\over d\eta_2}\right|_{\eta_2=\eta_1+\de}   \label {0}
 \eq
 involves a shift in $\eta_2$ of the charge hadron and an integration over the trigger $\eta_1$, and shows that the range of \cor\ in $\de$ is no more than the range of the inclusive \dis\ apart from the smearing of the trigger acceptance, which lengthens the $\Delta\eta$ range by 1.5. The implication is that there is no long-range longitudinal \cor. Any successful model of ridge formation should be able to explain the simple relationship shown in Eq.\ (\ref{0}).
 In Ref.\ \cite{ch1} an interpretation of that relationship is given in terms of transverse \cor\ that we discuss in more detail in the next section.
  
 \section{Ridge at $|\Delta\eta|>1$}
 
 The phenomenological verification of Eq.\ (\ref{0}) directs one's attention to the origin of ridge formation without intrinsic longitudinal \cor\ at large $\Delta\eta$. From all that have been learned experimentally about the ridges, there is no indication that such structure can be found in the absence of any jet. Even in autocorrelation studies where no triggers are used, ridges are found in the kinematical region where minijets are detected \cite{ja}. Our approach is therefore to start with jet-induced transverse \cor\ at $|\de|<1$ and to extend it to larger $\eta$ separation, in contrast to other studies where long-range longitudinal \cor\ at low $p_T$ exists without jets and then a large-$p_T$ parton is introduced to define the $\de$ range. The approach we adopt was actually advocated even before the discovery of ridge was reported by Putschke \cite{jp} at a time when the phenomenon was regarded as the pedestal lying under the jet peak \cite{ja1,ch2}. Now, with more data and model analyses of the transverse \cor\ at hand, the extension to large $\de$ can be done with more definiteness.
 
 To be more specific, let us consider the ridge found by CMS at LHC, where only charged particles with $|\eta|<2.4$ and $p_T<4$ GeV/c are used in the analysis. In that acceptance region the hadron
  $p_L$ is less than 22 GeV/c, so Feynman $x_F$ is $< 6.3 \times 10^{-3}$ at $\sqrt s=7$ TeV, and the corresponding partons that recombine have even lower $x$ values.  Those are soft wee partons deep in the sea, whose correlations can be strongly influenced by fluctuations.  
 Suppose that a semihard scattering occurs in a $pp$ collision at 7 TeV and sends a parton to the $\eta \approx 0$ region with a parton momentum $k_T$ in the 5-10 GeV/c range, which we shall regard as intermediate at LHC.  Whatever the medium effect on it may be, it can lead to a cluster of hadrons with limited range in $\eta$ and $\phi$ \cite{cms}.  It cannot directly cause the production of an associated particle at $\eta = 2.4$ since the $p_L$ of that particle can exceed 20 GeV/c, hence forbidden by energy conservation.  Any particle produced outside the jet peak  carries longitudinal momentum that is driven by the initial partons (right- or left-movers) of the incident protons.  In the conventional parton model it is assumed that there are no significant longitudinal constraints on those initial partons \cite{rf,bj}.  We add, however, that their transverse momentum distribution can be affected by the semihard scattering before they recede from one another.  At early time the right- and left-movers need not be arranged as in Hubble expansion, i.e., a right-moving parton may be located on the left side of the region of uncertainty, and vice-versa; hence, those initial partons can be sensitive the passage of the semi-hard parton across their ways.  The quantum fluctuations that generate the transverse $k_T$ distribution of the forward (or backward) moving partons may be enhanced by the energy loss of the semihard parton.  More specifically, let $\exp(-k_T/T)$ represent the $k_T$ distribution in the absence of semihard scattering; then our assertion is that the distribution changes to $\exp(-k_T/T')$ with $T' > T$ in the presence of semihard scattering, provided that the affected partons are in the vicinity of the semihard parton trajectory in the transverse plane, i.e., $\Delta\phi$ is limited on the near side.  Furthermore, such a change occurs for all partons independent of their longitudinal momenta up to $x \sim10^{-2}$, say.  
 This enhancement is in essence the transverse \cor\ discussed in Sec. II.A, but now the semihard parton at $\eta\approx 0$ induces a change in the transverse \dis\ of the soft partons from $T$ to $T'$ at all $\eta$  in the limited region $|\eta|<2.4$ under study.
 
 The CMS experiment does not identify any particle as the trigger, so the pseudorapidity of the semihard parton cannot be specified. All charged particles accepted in the window $|\eta|<2.4$ are used for the analysis of the two-particle \cor. Thus the correlated particles may be at $\eta_{1,2}=\pm 2.4$, resulting in $|\de|=|\eta_1-\eta_2|$ as large as 4.8. Hereafter, $\eta_1$ and $\eta_2$ do not refer to trigger and associated particles, respectively, but to any two particles whose correlation is measured by CMS. The semihard parton may be anywhere in between $\pm2.4$. The huge jet peak  observed in Ref.\ \cite{cms} corresponds to particles that are produced by thermal-shower recombination and therefore must be close in $\eta$ to the semihard parton, but the peak \dis\ in $\de$ does not indicate where it is. The flat ridge \dis\ that lies below the jet peak only reveals the response of the medium in terms of enhanced thermal partons without any information about the locations of the shower partons. The ridge particles have transverse distribution that is characterized by the same inverse slope $T'$ as for the enhanced soft partons. That is a property of recombination \cite{hy,hz1}. No explicit longitudinal \cor\ has been put in. 
 
 In order to describe pion and proton production in the same formalism of recombination of thermal partons at low $p_T$, it is shown that the replacement of $p_T$ by $E_T$, where $E(p_T)=(m_h^2+p_T^2)^{1/2}-m_h, h=\pi$ or $p$, is sufficient to account for the mass effect and that the inclusive ridge \dis\ can reproduce $v_2^h(E_T,N_{part})$ at low $E_T$ \cite{hz1}. Being the difference between the enhanced \dis\ and the background, that ridge \dis\ is 
 \bq
 R(p_T)=R_0(e^{-E_T/T'}-e^{-E_T/T})   \label{1a}
 \eq
 for nuclear collisions. It is the soft response to the semihard partons. 
  We will apply the same description to $pp$ collision below. The difference $\Delta T=T'-T$ is a measure of the magnitude of the influence by semihard scattering without which there is no ridge.

 \section{Ridge Yield in $pp$ Collision at LHC}
 
 We now focus on the ridge yield measured by CMS.
Let the single-particle distribution be $\rho(p_T,\eta) = dN/p_Td\eta dp_T$, which will be abbreviated by $\rho_1(i)$ for the $i$th particle, so that two-particle distribution is denoted by $\rho_2(1,2)$.  Define two-particle correlation by 
$C_2(1,2) = \rho_2(1,2)-\rho_1(1)\rho_1(2).$
  The measure for ridge used by CMS is
\begin{eqnarray}
{\cal R}_{\rm CMS}(1,2) = NC_2(1,2)/\rho_1(1)\rho_1(2),     \label 1
\end{eqnarray}
where $N$ is the number of charged particles in a multiplicity bin.  In more detail the quantities in Eq.\ (\ref{1}) are averaged over bins of $p_T$, so Ref.\ \cite{cms} exhibits
\begin{eqnarray}
{\cal R}_{\rm CMS}(p_T,\Delta\eta,\Delta\phi) = N{\prod_{i=1,2}\left[\int_{[p_T]}dp_{Ti}p_{Ti}\right] C_2(1,2) \over \prod_{i=1,2}\left[\int_{[p_T]}dp_{Ti}p_{Ti} \rho_1(i)\right]}     \label 2
\end{eqnarray}
where $[p_T]$ denotes the range of integration from $p_T -0.5$  to $p_T +0.5$ (GeV/c).  A ridge then appears in the 2D $\Delta\eta$-$\Delta\phi$ distribution.  A projection of it onto $\Delta\phi$ is done by integrating $|\Delta\eta|$ over the range 2.0 to 4.8.  The associated yield in the ridge is then determined by integrating over a range of $\Delta\phi$ around 0 where ${\cal R}_{\rm CMS}$ is above its minimum, i.e., 
\begin{eqnarray}
Y_R(p_T,N)= \int_R d\Delta\phi \int_{\pm2}^{\pm4.8} d\Delta\eta\ {\cal R}_{\rm CMS}(p_T,\Delta\eta,\Delta\phi).     \label 3
\end{eqnarray}
This measure of the ridge yield is given for 4 bins of $p_T$ and $N$ each \cite{cms}.  The data points are shown in Fig.\ 1.

\begin{figure}[tbph]
\hspace*{-.6cm}
\includegraphics[width=0.4\textwidth]{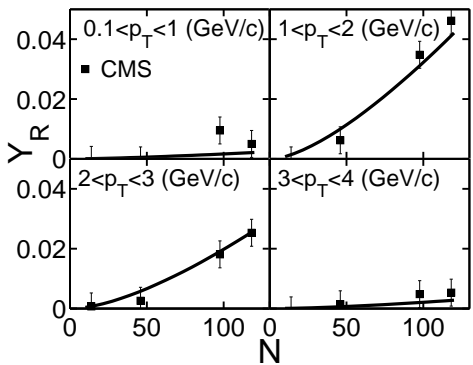}
\vspace*{-.5cm}
\caption{Ridge yield vs multiplicity $N$ for 4 bins of \pt. Data are from Ref.\ \cite{cms}, and lines are from model calculation.}
\end{figure}

What is remarkable about the data is that $Y_R$ is very small for both $0.1 < p_T <1$ and $3 < p_T < 4$ GeV/c, but jumps up by nearly an order of magnitude in the $1 < p_T < 2$ GeV/c bin.  It is very unusual in high-energy physics where the $p_T$ behavior is so drastically different on the two sides of 1 GeV/c.  The increase of $Y_R$ with $N$ is not surprising, especially if one has in mind that jets are connected with the ridge phenomenon.

Our explanation of the $p_T$ and $N$ dependencies of $Y_R$ is very simple, based on what has already been discussed.  We assume no longitudinal \cor, as in \cite{ch1}, which can explain the PHOBOS data \cite{ba}. Thus the only contribution to $C_2(1,2)$ is from transverse \cor\ that gives rise to the ridge \dis\ given in Eq.\ (\ref{1a}) as an $\eta$-independent response to the semihard jet at any $\eta_{jet}$. We therefore write
\bq
C_2(1,2)=R(1)R(2).   \label{2a}
\eq
This is a very unconventional description of correlation that we are proposing, since one usually expects an unfactorizable form for correlation.  The two particles at $\eta_1$ and $\eta_2$ are correlated because their $p_T$ distributions are both enhanced by the jet. $R(1)$ and $R(2)$ are independent responses, so they enter into $C_2(1,2)$ as factorized products. 
We emphasize that Eq.\ (\ref{2a}) is a correlation between two soft particles, each of which being correlated transversely to the unobserved jet as described by Eq.\ (\ref{1a}).  An analogy for this is the adage that  rising tide raises all boats --- even though, we add, there are no intrinsic horizontal correlations among the boats.
Putting Eq.\ (\ref{2a}) in (\ref{2}) and (\ref{3}) we obtain
\begin{eqnarray}
Y_R(p_T,N) = cN \prod_{i=1}^2\left[{\int_{[p_T]}dp_{Ti}p_{Ti}R(p_{Ti},N) \over \int_{[p_T]}dp_{Ti}p_{Ti} \rho_1(p_{Ti})}\right] ,    \label 5
\end{eqnarray}
where $c$ is an adjustable parameter that depends on the experiment.
This is an explicit formula that enables us to do phenomenological analysis.

The single-particle distribution for $|\eta | < 2.4$ at 7 TeV is given by CMS in the Tsallis parametrization \cite{vk}
\begin{eqnarray}
\rho_1(p_T) = \rho_0(1 + {E_T\over nT_0})^{-n}  \label 6
\end{eqnarray}
with $T_0 = 0.145$ GeV/c and $n = 6.6$.  The average $p_T$ found from the above fit is $\langle p_T \rangle = 0.545$ GeV/c. 

We use Eq.\ (\ref{6}) in (\ref{5}) and fit the data in Fig.\ 1 with two parameters (apart from normalization), which we choose to be $T$ and $\beta$, where
\begin{eqnarray}
{\Delta T\over T} = \beta \ln N,  \qquad  \Delta T = T' - T.   \label 7
\end{eqnarray}
This dependence on $N$ is reasonable, since at higher $N$ there is higher probability for jet production and hence larger $\Delta T$, which is in the exponent in Eq.\ (\ref{1a}).  The result of the fit is shown by the solid lines in Fig.\ 1 for
\begin{eqnarray}
T = 0.294\  {\rm GeV} \qquad {\rm and} \qquad  \beta = 0.0175.  \label 8
\end{eqnarray}
Evidently, our model reproduces the data very well for all $p_T$ and $N$ bins.  $Y_R(p_T,N)$ is small at small $p_T$ because $R(p_T)$ in Eq.\ (\ref{1a}) is suppressed as $p_T \rightarrow 0$. The reason for that is discussed below. $Y_R(p_T, N)$ is also small at large $p_T$; that is due both to the exponential suppression of $R(p_T)$ and the power-law decrease of $\rho_1(p_T)$ at high $p_T$. The increase with $N$ that is most pronounced in the $1<p_T<2$ GeV/c bin, where $R(p_T)$ is maximum, is clearly due to the enhancement of $T$ when jet production is more likely in accordance to Eq.\ (\ref{7}). At $N=100$, $\Delta T/T$ is about 8\%, which is slightly lower than that observed in nuclear collisions at RHIC where $T=355\pm 6$ MeV/c and $T'=416\pm 22$ MeV/c for $4<p_T^{\rm trig}<6$ GeV/c \cite{bia}.

The reason why $R(p_T)$ must vanish as $p_T \to 0$ is related to azimuthal anisotropy in nuclear collisions. We have advocated the view that the ridge component before being averaged over $\phi$ contains all the $\phi$ dependence of the inclusive distribution \cite{chy, rch}. In that approach which has been worked out in more detail recently in \cite{hz1}, it is shown without using hydrodynamics that $v_2$ (referred to as elliptic flow in hydro description)  can be reproduced at all centralities, provided that $R(p_T)\to 0$ at vanishing $p_T$ because $v_2(p_T)\to 0$. Since the azimuthal behavior is determined primarily by the initial geometry of the collision system \cite{chy,hz,rch}, such an approach may well be applicable to $pp$ collisions, for which the validity of hydrodynamics used for nuclear collisions is doubtful. The origin of the $\phi$ dependence in the geometrical approach is the anisotropy of semihard emission when the initial configuration is almond-shaped.  Similarly, it is reasonable to consider the initial configuration in $pp$ collisions also, when the impact parameter is non-zero, and we expect  significant $\phi$ anisotropy in the produced particles.

	The Tsallis \dis\ in Eq.\ (\ref{6}) has the property of a power-law behavior at large \pt, but an exponential behavior, $\exp(-E_T/T_0)$, at low \pt. It is then of interest to note the difference between the values of $T_0$ and $T$, the latter being twice larger than the former. It may appear as being inconsistent; however, the average $\left< p_T \right>$ of $\exp(-E_T/T)$ is 0.6 GeV/c, only 10\% higher than that for Eq.\ (\ref{6}). Thus different parametrizations of the $E_T$ \dis\ give essentially the same physical quantity. Eq.\ (\ref{6}) is a fit of the CMS data \cite{vk} that emphasizes the $p_T^{-n}$ behavior at high \pt, while Eq.\ (\ref{1a}) is a theoretical model of the ridge \dis\ at low \pt.

\section{Conclusion}

	We have given an interpretation of the ridge phenomenon in $pp$ collisions in terms of soft partons on which very little is known. By drawing on what we do know about the soft partons in nuclear collisions, we are led to the implication that  a dense medium can be created even in $pp$ collisions at 7 TeV. The primary input in our approach to explaining the observed ridge yield is the assertion that the correlation is of the factorizable form $R(1)R(2)$, where $R(i)$ is the response of the {\it i}th soft particle to the unobserved jet, so that two independent transverse \cor s of the semihard-soft type can lead to a net soft-soft \cor\ in $C_2(1,2)$.
	
	The success of our approach applied to $pp$ collisions at 7 TeV suggests that (a) the medium can be responsive to semihard jets, (b) there can be azimuthal anisotropy, (c) the \pt\ spectrum in the ridge is harder than that of the inclusive, and (d) that hadronization is by recombination. None of the above rely on the validity of hydrodynamics for $pp$ collisions, or the existence of intrinsic long-range longitudinal \cor, and all of them can be checked by further experimental measurements. The last item cannot be checked directly, but one of its consequences is that the $p/\pi$ ratio can be large,  which is a property of all recombination/coalescence models \cite{reco}. We expect  the $p/\pi$ ratio in the ridge to increase with \pt\ at low \pt\ in $pp$ collisions at 7 TeV, although the rate of that increase depends on the soft parton density, on which we have insufficient knowledge to predict.  A ratio larger than 0.2 cannot be explained by fragmentation. Thus the experimental determination of the $p/\pi$ ratio in the ridge will be very interesting and should provide further insight on the structure and origin of the ridge.

The basic issue that the observation of a ridge by CMS has opened up is whether a system of high density soft partons can be created in $pp$ collisions. The system may be too small for the applicability of hydrodynamics, but azimuthal anisotropy can nevertheless exist for small systems in non-central collisions, so consequences on $\phi$ asymmetry should be measurable, as the ridge structure on the near side demonstrates. Our consideration of ridge formation as being generated by semihard jets applies to both hadronic and nuclear collisions. Thus we go further to suggest that even in single-particle \dis\ in $pp$ collisions at LHC there may exist a ridge component that contains all the $\phi$ dependence, as found in Au-Au collisions \cite{chy,hz,hz1}.


This work was supported  in
part,  by the U.\ S.\ Department of Energy under Grant No. DE-FG02-96ER40972 and by the National Natural Science Foundation of China under Grant No.\ 10775057 and 11075061.

\end{document}